\documentclass[twocolumn,amsmath,amssymb,showpacs]{revtex4}
\usepackage{graphicx}

\usepackage{longtable}
\usepackage{times}
\usepackage{dcolumn}
\usepackage{bm}
\usepackage[plainpages = false, pdfpagelabels, 
bookmarks,
bookmarksopen = true,
bookmarksnumbered = true,
linktocpage,
pagebackref=false,
colorlinks = true,
linkcolor = blue,
urlcolor  = blue,
citecolor = blue,
anchorcolor = green,
hyperindex = true,
hyperfigures]
{hyperref} 

\begin{document}

\title {\Large Neutron radii and semi-phenomenological treatment of neutron distributions for Mg isotopes }

\author{Govind \surname{Kumar}$^1$}
\author{M. \surname{Imran}$^{2,3}$}
\author{Z. \surname{Hasan}$^1$}
\author{Z. A. \surname{Khan}$^2$}\email{zakhan.amu@gmail.com}
\affiliation{$^1$Department of Applied Physics, ZH College of Engineering and Technology, Aligarh Muslim University, Aligarh-202002, India}
\affiliation{$^2$Department of Physics, Aligarh Muslim University, Aligarh-202002, India}
\affiliation{$^3$Applied Science \& Humanities Section, University Women's Polytechnic, Aligarh Muslim University, Aligarh-202002, India}

\begin{abstract}

Involving the charge radii of \rm Mg isotopes, as calculated using the deformed relativistic Hartree-Bogoliubov theory in continuum (DRHBc), we have extracted the neutron radii of $^{24-38}$\rm Mg isotopes by studying their reaction cross sections ($\sigma_{R}$) from $^{12}$\rm C at 240 MeV/nucleon within the framework of Glauber model. The calculations use (i) descriptions of nuclei in terms of the Slater determinant involving harmonic oscillator single-particle wave functions (SDHO), and (ii) two-parameter Fermi (2pF) shape of density distribution, with the aim to assess the density dependence of neutron skin in $^{24-38}$\rm Mg isotopes. To understand the asymptotic behavior (spread) of neutron distribution, we propose to introduce the use of core+n ($S_{n}<S_{2n}$) or core+2n ($S_{n}>S_{2n}$) description for stable as well as unstable isotopes; $S_{n}$ ($S_{2n}$) is the one-neutron (two-neutron) separation energy of the considered isotope. The core+n (core+2n) is treated semi-phenomenologically. In this work, the core+n is employed for $^{25-38}$\rm Mg isotopes, and is subjected to reproduce the same neutron radius of the given isotope, as we obtained from $\sigma_{R}$ calculations. To validate the core+n description, we have revisited the reaction cross sections of $^{25-38}$\rm Mg isotopes. The results are found to agree well with the experimental values. Moreover, the core+n neutron distributions clearly demonstrate the one-neutron halo structure of $^{37}$\rm Mg. These findings motivated us to use the core+2n description for the neutron distribution of $^{40}$\rm Mg in predicting its neutron radius, and $\sigma_{R}$ from $^{12}$\rm C at 240 and 1000 MeV/nucleon. The trend of the neutron radius and $\sigma_{R}$ suggests that $^{40}$\rm Mg exhibits two-neutron halo like structure.

\end{abstract}
\maketitle

\section{Introduction}
\label{sec1}

The reliable information about the distribution of protons and neutrons plays an essential role to understanding the occurrence of some exotic phenomena in neutron-rich (unstable) nuclei, particularly the neutron halos and neutron skin, and the equation of state (EOS) of nuclear matter that may provide the properties of astronomical objects. The quantity connecting the nucleon distribution is the nuclear size which is understood in terms of the root-mean-square (rms) proton (charge) and neutron radii. Among the available methods used to determine the nuclear size, the electron scattering and isotope shift measurements \cite{1,2} have been performed to predict charge radii covering a large section of stable and a limited number of unstable (short lived) nuclei around the $\beta$-stability line; the proton elastic scattering and the reaction (interaction) cross sections have been quite successful to extract the neutron radii of such nuclei. However, the advancement of Radioactive Ion Beam (RIB) facility allowed to produce unstable nuclei far away from the stability line, thus facing a challenging task to study nuclear sizes up to the most exotic nuclei. To accommodate the exotic (unstable) nuclei, the measurement of charge-changing cross section (CCCS), describing the probability of a projectile nucleus losing its proton(s) by interacting with the target, has been performed that is considered as an alternative to the electron scattering experiments to investigating the proton distribution in nuclei \cite{3,4,5,6,7,8,9}. More explicitly, it was thought that projectile protons may account for the charge-changing cross section data. However, the calculations of CCCS within the framework of the Glauber model \cite{4,10,11,12,13,14,15} have shown that the projectile neutrons may partially contribute to the CCCS, giving rise a complex understanding of the reaction mechanism needed to study CCCS. Keeping in view the status of the reaction mechanism for CCCS, it has been suggested that the study of CCCSs could be possible, involving only the projectile protons, provided one introduces a phenomenological scaling (correction) parameter \cite{4,10,11,12,13,16,17,18} that accommodates the contribution due to the presence of neutrons in the projectile. The results of such calculations \cite{16} are found to provide useful information about the proton (charge) radii of the exotic isotopes of relatively lighter elements Z $\leq$ 9, and the combined study of CCCSs and interaction cross sections is expected to give reliable estimates for the neutron skin thickness and understanding the halo-like structures of neutron rich (unstable) nuclei \cite{16}. 

Despite reasonable success in predicting the proton radii of exotic nuclei through CCCS data, the isotopic chains of the elements \rm Mg, \rm Ne, and \rm Na lack in the  measurements on their charge-changing cross sections. In such cases, one, therefore, relies on some model-dependent estimates for the charge radii, and involves the analysis of their reaction (interaction) cross sections to providing information about the neutron skins. In this work, we consider one such element, say \rm Mg, and study the reaction cross sections ($\sigma_{R}$) of its isotopes on a $^{12}$\rm C target at 240 MeV/nucleon \cite{19} in the framework of the Glauber model. The aim of this study is to extract the neutron radii of $^{24-38}$\rm Mg isotopes, taking the proton radii obtained from the deformed relativistic Hartree-Bogoliubov theory in continuum (DRHBc) calculations \cite{20}. The calculations involve descriptions of nuclei in terms of the Slater determinant consisting of the  harmonic oscillator single-particle wave functions (referred to as SDHO density) and the two-parameter Fermi (2pF) density. The idea is to assess how far the choice of different forms of density distribution affects the extracted neutron skin thickness. In the next part of our calculations, we proceed to study the asymptotic behavior (spread) of neutron distributions in $^{25-38}$\rm Mg isotopes. For this, we have used the core+n description for a given ($N>Z$) \rm Mg isotope; the core+n is restricted to reproduce the same neutron radius, as we obtained from the analysis of reaction cross sections. Finally, using the core+2n description for the neutron distribution of $^{40}$\rm Mg, we have predicted its neutron radius, and $\sigma_{R}$ from $^{12}$\rm C at 240 and 1000 MeV/nucleon. The extracted values of neutron radius and $\sigma_{R}$ support the two-neutron halo structure of $^{40}$\rm Mg \cite{21}.   
                          
The formulation of the problem is given in  Sec. \ref{sec2}. The numerical results are presented and discussed in Sec. \ref{sec3}. The conclusions are presented in Sec. \ref{sec4}.

\section{Formulation}

\label{sec2}

According to the Glauber model, the reaction cross section ($\sigma_{R}$) for the scattering of a projectile nucleus with a ground-state wave function $\psi_{P}$ on a target nucleus with a ground-state wave function $\psi_{T}$ is given by
\begin{equation}
\sigma_{R}=2\pi \int [1-|S_{el}(b)|^{2}]bdb,
\label {eq1}
\end{equation}
\begin{equation}
	S_{el}(b)= \langle \psi_{T}\psi_{P}\vert\prod^{A}_{i=1}\prod^{B}_{j=1}[1-\Gamma_{NN}(\vec{b}-\vec{s_{i}}+\vec{s^{'}_{j}})]\vert\psi_{P}\psi_{T}\rangle,
	\label{eq2}
\end{equation}
where $A(B)$ is the mass number of target (projectile) nucleus, $\vec{b}$ is the impact parameter vector perpendicular to the incident momentum, $\vec {s_{i}}\vec{(s^{'}_{j}})$ are the projections of
target (projectile) nucleon coordinates on the impact parameter plane, 
and $\Gamma_{NN}({b})$ is the $NN$ profile function, which is related to the $NN$ scattering amplitude $f_{NN}(q)$ as 
follows  
\begin{equation}
	\Gamma_{NN}({b})= \frac{1}{2\pi ik}~\int \exp(-i\vec {q}.\vec {b})f_{NN}({q})~d^{2}q, 
	\label{eq3}
\end{equation}
where $k$ is the incident nucleon momentum corresponding to the projectile kinetic energy per nucleon, and $\vec{q}$ is the momentum transfer.

Following the approach of Ahmad \cite{22}, the S-matrix element, $S_{el}({b})$, up to two-body correlation (density) term takes the following form:
\begin{equation}
	S_{el}({b})\approx S_{0}(b)+S_{2}(b),
	\label {eq4}
\end{equation}
where
\begin{equation}
	S_{0}({b})= [1-\Gamma^{NN}_{00}(b)]^{AB}, 
	\label{eq5}
\end{equation}
and
\begin{gather}
	S_{2}({b})=\Bigl\langle\psi_{T}\psi_{P}\Bigl|\frac{1}{2!}[1-\Gamma^{NN}_{00}(b)]^{AB-2}~~~~~~~~~~~~~~~~~~\nonumber\\
	~~~~~~~~~~~~~~~~~~\times \sum^{'}_{i_{1},j_{1}}\sum^{'}_{i_{2},j_{2}}\gamma_{i_{1},j_{1}}\gamma_{i_{2},j_{2}}\Bigr|\psi_{P}\psi_{T}\Bigr\rangle,
	\label{eq6}
\end{gather}
with 
\begin{equation}
	\gamma_{ij}=\Gamma^{NN}_{00}(\vec{b})-\Gamma_{NN}(\vec{b}-\vec{s_{i}}+\vec{s^{'}_{j}}), 
	\label{eq7}
\end{equation}
and
\begin{equation}
	\Gamma^{NN}_{00}({b})=\int \rho_{T}(\vec r)~\rho_{P}(\vec{r^{'}})~\Gamma_{NN}(\vec b-\vec s +\vec{s^{'}})~d\vec r~d\vec{r^{'}}.
	\label {eq8}
\end{equation}
The primes on the summation signs in Eq. (\ref{eq6}) indicate the restriction that two pairs of indices can not be equal at the same time (for example,
if $i_{1}$ = $i_{2}$ then $j_{1} \neq j_{2}$ and vice versa). The quantities $\rho_{T}$ and $\rho_{P}$ in Eq. (\ref{eq8}) are the (one-body) ground state
densities of the target and projectile, respectively. From calculations point of view, it should be noted that the distinction between protons and neutrons in both the projectile and target has been included in the uncorrelated part ($S_{0}$) of the S-matrix element only. This makes the use of different values of the parameters for pp and pn scattering amplitudes and considers different density distributions for protons and neutrons in the colliding nuclei. However, it turns out that inclusion of the distinct features of protons and neutrons in the two-body correlation term ($S_{2}({b})$) is not as straightforward as it was in $S_{0}({b})$. Therefore, the two-body correlation term uses the average behavior of pp and pn interactions, and involves matter density distributions. More explicitly, the evaluation of $S_{0}({b})$ and $S_{2}({b})$ leads to the following expressions:
\begin{gather}
	S_{0}({b})= [1-\Gamma^{pp}_{00}({b})]^{Z_{P}Z_{T}}[1-\Gamma^{np}_{00}({b})]^{N_{P}Z_{T}}~~~\nonumber\\ 
	~~~~~~~~~~~~~~~~~\times[1-\Gamma^{pn}_{00}({b})]^{Z_{P}N_{T}}[1-\Gamma^{nn}_{00}({b})]^{N_{P}N_{T}},
	\label{eq9}
\end{gather}
\begin{gather}
	S_{2}({b})=-\frac{AB}{8\pi^{2}k^{2}}[1-\Gamma^{NN}_{00}(b)]^{AB-2}[(A-1)(B-1)\nonumber\\
	~~~~~~~~~~~~~~~~~\times(G_{22}(b)-G_{00}^{2}(b))+(B-1)\nonumber\\
	~~~~~~~~~~~~~~~~~\times(G_{21}(b)-G_{00}^{2}(b))+(A-1)\nonumber\\
	~~~~~~\times(G_{12}(b)-G_{00}^{2}(b))],
	\label{eq10}
\end{gather}
with
\begin{equation}
	\Gamma^{ij}_{00}({b})=\int \rho^{j}_{T}(\vec r_{j})~\rho^{i}_{P}(\vec{r_{i}^{'}})~\Gamma_{ij}(\vec b-\vec s_{j} +\vec{s_{i}^{'}})~d\vec r_{j}~d\vec{r_{i}^{'}},
	\label{eq11}
\end{equation}
where $Z_{T}(Z_{P})$ and $N_{T}(N_{P})$ are the target (projectile) atomic and neutron number, respectively, and each of $i$ and $j$ stands for a proton and a neutron. For details of $G_{22}(b), G_{21}(b), G_{12}(b)$, and  $G_{00}(b)$, we refer readers to follow the work presented in Ref. \cite{16}.

\section{Results and Discussion}
\label{sec3}

Following the approach outlined in Sec. \ref{sec2}, we have analyzed the experimental reaction cross sections of $^{24-38}$\rm Mg isotopes on a $^{12}$\rm C target at 240 MeV/nucleon \cite{19}, and predicted matter radii of the projectiles. We have also predicted the matter radius of $^{40}$\rm Mg, and its reaction cross section from $^{12}$\rm C at 240 and 1000 MeV/nucleon.
The inputs required in the calculation are (i) the nucleon-nucleon (NN) scattering amplitude, and (ii) the proton and neutron density distributions of colliding nuclei. 

(i) NN scattering amplitude: The NN scattering amplitude $f_{NN}(q)$ is usually parametrized in the form \cite{23}
\begin{equation}
f_{NN}(\vec q)=\frac{ik\sigma_{NN}}{4\pi}
(1-i\rho_{NN})\exp\left(-\beta_{NN}q^{2}/2\right),
\label{eq12}
\end{equation}
where $\sigma_{NN}$ is the NN total cross section, $\rho_{NN}$ the ratio of the real to the imaginary parts of the forward NN amplitude, and $\beta_{NN}$ is the slope parameter. In the present work, we need the values of $\sigma_{NN}$, $\rho_{NN}$, and $\beta_{NN}$ at 240 and 1000 MeV, which are taken from Ref. \cite{23}. 

(ii) Matter density distributions of the colliding nuclei: For the target $^{12}$\rm C nucleus, we use the charge density as obtained from the electron scattering experiment \cite{1} and assume the neutron and proton densities to be the same.  
The intrinsic proton and neutron densities of the projectile, used in this work, have been taken in the following forms:

SDHO density: The SDHO density has been obtained from the Slater determinant consisting of the harmonic oscillator single-particle wave functions \cite{24}. For completeness of the discussion, let us highlight the important steps, needed to calculate the SDHO density. The first step is to calculate the nucleon density which accounts the effect due to centre-of-mass (cm) motion
\begin{equation}
\bar{\rho}(\vec{r})= \langle\Psi\vert\sum^{M}_{i=1}\delta(\vec{r}_{i}-\vec{r})P_{i}\vert\Psi\rangle,
\label {eq13}
\end{equation}
where $\Psi$ is the Slater determinant, $M$ is the number of nucleons in the nucleus, $\vec{r}_{i}$ is the nucleon coordinate, and $P_{i}$ is a projection to proton/neutron. Involvement of the single-particle harmonic oscillator wave functions leads to the following expression for $\bar{\rho}$:
\begin{equation}
\bar{\rho}(\vec{r})=\frac{1}{4\pi}\sum_{nlj}N_{nlj}R^{2}_{nlj}(\vec{r}),
\label {eq14}
\end{equation}
where $N_{nlj}$ and $R_{nlj}$ denote the occupation number and the radial wave function of the $nlj$ orbit, respectively.

To obtain the intrinsic nucleon density which is free from the cm effects, one uses the fact that the cm motion $\Psi_{cm}(\vec{X})$ contained in $\Psi$ is factorized as 
\begin{equation}
\Psi=\Psi_{0}\Psi_{cm}(\vec{X}),
\label {eq15}
\end{equation}
where $\vec{X}$ is the cm coordinate and
\begin{equation}
\Psi_{cm}(\vec{X})=\left(\frac{M\alpha^{2}}{\pi}\right)^{3/4}\exp\left(-\frac{M\alpha^{2}}{2}\vec{X}^{2}\right).
\label {eq16}
\end{equation}
With the above considerations, the intrinsic nucleon density $\rho(\vec{r})$,
\begin{equation}
\rho(\vec{r})=\langle\Psi_{0}\vert\sum^{N}_{i=1}\delta(\vec{r}_{i}-\vec{X}-\vec{r})P_{i}\vert\Psi_{0}\rangle,
\label {eq17}
\end{equation}
can be obtained from the inverse Fourier transform of the following relation \cite{24}:
\begin{equation}
\int d\vec{r}e^{i\vec{k}.\vec{r}}\rho(\vec{r})=\left[\langle\Psi_{cm}\vert e^{i\vec{k}.\vec{X}}\vert\Psi_{cm}\rangle\right]^{-1}\int d\vec{r}e^{i\vec{k}.\vec{r}}\bar{\rho}(\vec{r}), 
\label {eq18}
\end{equation}
with
\begin{equation}
\langle\Psi_{cm}\vert e^{i\vec{k}.\vec{X}}\vert\Psi_{cm}\rangle=\exp \left(-\frac{k^{2}}{4M\alpha^{2}}\right).
\label {eq19}
\end{equation}
The expressions for the intrinsic proton and neutron density distributions of the projectile nuclei, as obtained above, are taken from Ref. \cite{24}. These density distributions are, hereafter, referred to as SDHO densities. They involve the oscillator constant $\alpha^{2}$ as their basic input, which assumes different values for proton ($\alpha^{2}_{p}$) and neutron ($\alpha^{2}_{n}$) density distributions, providing the rms proton and neutron radii of a nucleus under consideration.  
  
Two-parameter Fermi (2pF) density: The functional form of two-parameter Fermi density distribution is given by
\begin{equation}
	\rho(r)=\frac{\rho_{0}}{1+e^{(r-R)/a}},
	\label{eq20}
\end{equation}
where $a$ is the diffuseness parameter and $R$ = $r_{0}Z^{1/3}$(or$N^{1/3}$) \cite{25} ($r_{0}$=1.2 fm) is the radius parameter. The central density $\rho_{0}$ is determined by the normalization to the number of protons ($Z$) or neutrons ($N$). It is, thus, obvious that, in the 2pF distribution, the diffuseness parameter $a$ is the only variable parameter, which is responsible to provide the rms proton and neutron radii of a given nucleus. For further discussion, we represent $a_{p}$ and $a_{n}$ as the diffuseness parameters for proton and neutron distributions, respectively.

In order to appreciate the context of the proposed work, we present the results of our calculations in the following manner. 
  
\subsection{Predictions for reaction cross sections of $\textsuperscript{24-38}\rm \text{Mg}$ isotopes on $\textsuperscript{12}\rm \text{C}$  at 240 MeV/nucleon, involving the DRHBc proton and neutron radii \cite{20}}
\begin{figure}
	\begin{center}
		\includegraphics[height=6.9cm, width=8.9cm]{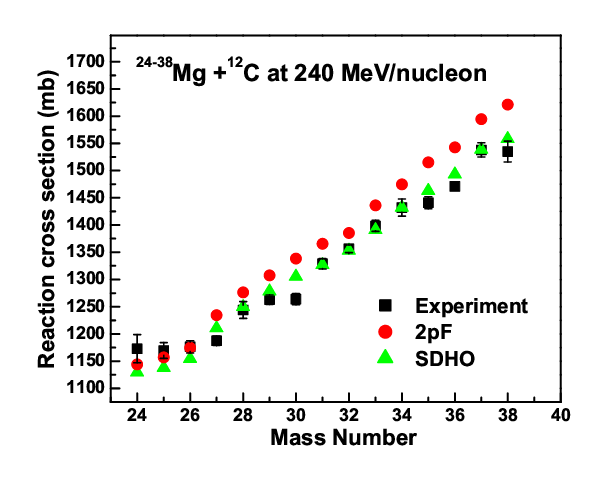}
		\caption{The reaction cross sections for \rm Mg isotopes on a $^{12}$\rm C target at 240 MeV/nucleon, using the proton and neutron radii, as calculated using the deformed relativistic Hartree-Bogoliubov theory in continuum (DRHBc) \cite{20}. Filled circles and triangles correspond, respectively, to 2pF and SDHO densities. The experimental data (filled squares) are taken from Ref. \cite{19}. }  
		\label{fig1}
	\end{center}
\end{figure}

\begin{figure}
	\begin{center}
		\includegraphics[height=8.9cm, width=8.9cm]{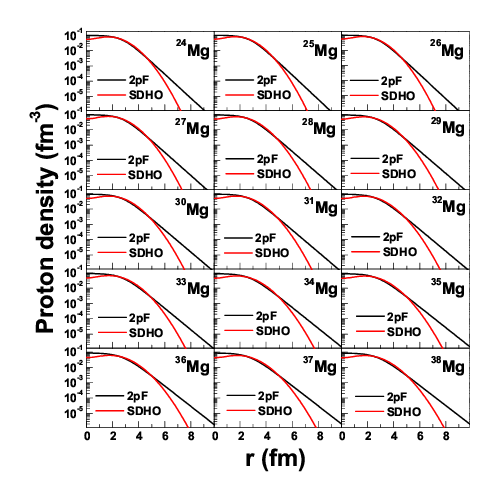}
		\caption{The 2pF (black line) and SDHO (red line) proton density distributions in $^{24-38}$\rm Mg isotopes, obtained using the DRHBc proton radii \cite{20}.}  
		\label{fig2}
	\end{center}
\end{figure}

\begin{figure}
 	\begin{center}
 		\includegraphics[height=8.9cm, width=8.9cm]{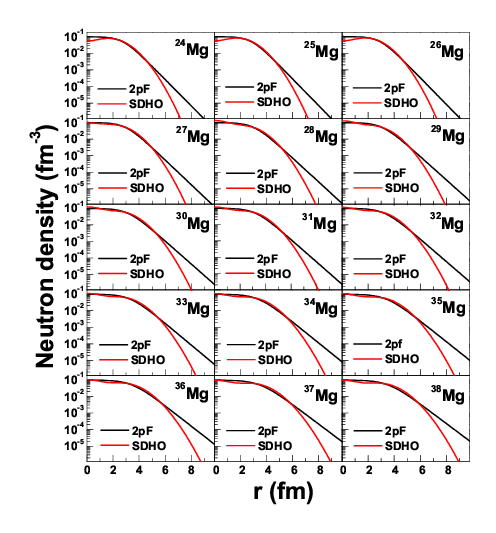}
 		\caption{The 2pF (black line) and SDHO (red line) neutron density distributions in $^{24-38}$\rm Mg isotopes, obtained using the DRHBc neutron radii \cite{20}.}  
 		\label{fig3}
 	\end{center}
 \end{figure}

The first part of the study deals with the calculation of reaction cross sections for $^{24-38}$\rm Mg isotopes on a $^{12}$\rm C target at 240 MeV/nucleon, involving the proton and neutron radii obtained in DRHBc calculations \cite{20}. For this, the oscillator constant ($\rm \alpha^2$) and the diffuseness parameter ($a$) for proton and neutron distributions are fixed from the corresponding DRHBc radii. The values of $\rm \alpha^2$ and $a$, along with the DRHBc proton and neutron radii are given in Table \ref{tab1}, and the results of the calculation for reaction cross section are presented in Fig. \ref{fig1}. It is found that in most of the cases, the predictions for $\sigma_{R}$ deviate from the experimental values; the deviation using the SDHO density is found to be less as compared to the 2pF density, suggesting that SDHO density could be a better choice to describe reaction cross sections. To understand the plausible cause for the difference between SDHO and 2pF results for $\sigma_{R}$, we have plotted the corresponding proton and neutron distributions in Figs. \ref{fig2} and \ref{fig3}. In order to correlate our $\sigma_{R}$ results (Fig. \ref{fig1}), using SDHO and 2pF densities, with the density distributions (Figs. \ref{fig2} and \ref{fig3}), we proceed as follows. In the works of Aumann \textit{et al.} \cite{26,27} on knock-out reactions, and Imran \textit{et al.} \cite{16} on interaction cross sections, we have noticed that a complex target is sensitive to the surface region of the projectile, unlike the proton target which can explore both the inner and surface regions of the projectile. Keeping this in view, one may say that the deviation in the SDHO and 2pF surface densities could be the main source for different estimates of  $\sigma_{R}$ (Fig. \ref{fig1}).
\begin{table*}
	\caption{The diffuseness parameter for 2pF proton (neutron) density distribution, $a_{p}$ ($a_{n}$), and the oscillator constant for SDHO proton (neutron) density distribution, $\rm \alpha_p^2$ ($\rm \alpha_n^2$), reproduce the projectile proton; $r_{p}$ (neutron; $r_{n}$) radius as calculated using the deformed relativistic Hartree-Bogoliubov theory in continuum (DRHBc) (given in the last two columns). Both $r_{p}$ and $r_{n}$ take care of the finite size of the nucleon.}
	\renewcommand{\tabcolsep}{0.14mm}
	\renewcommand{\arraystretch}{1.2}
	\label{tab1}
	\begin{ruledtabular}
		\begin{tabular}{ccccccccccccc}
			Projectile & \multicolumn{2}{c}{$\rm 2pF$} &\multicolumn{2}{c}{$\rm SDHO $}&\multicolumn{2}{c}{$\rm DRHBc$ \cite{20}}\\
			\cline{2-3} \cline{4-5} \cline{6-7}
			& $a_{p}(fm)$  &$a_{n}(fm) $ & $\alpha^{2}_{p}(fm^{-2})$ &$\alpha^{2}_{n}(fm^{-2}) $ &$r_{p}(fm)$ &$r_{n}(fm) $  \\
			\hline
			$\rm ^{24}Mg$   &0.5567&0.5478&0.2956&0.3003 &2.968  &2.945                                       \\
			$\rm ^{25}Mg$   &0.5449&0.5406&0.3022&0.3030 &2.937  &2.969                          \\
			$\rm ^{26}Mg$   &0.5408 &0.5425&0.3045&0.3003 &2.927  &3.014                            \\
			$\rm ^{27}Mg$   &0.5752 &0.5836&0.2869&0.2789 &3.016  &3.156                            \\
			$\rm ^{28}Mg$   &0.5910 &0.6035&0.2794&0.2682 &3.058  &3.243                            \\  
			$\rm ^{29}Mg$   &0.5962 &0.6180&0.2771&0.2603 &3.072  &3.315                            \\
			$ \rm ^{30}Mg$   &0.6030 &0.6286&0.2741&0.2541 &3.090  &3.375                            \\
			$\rm ^{31}Mg$   &0.6025 &0.6335&0.2744&0.2501 &3.089  &3.419                            \\
			$\rm ^{32}Mg$   &0.6118 &0.6415&0.2702&0.2451 &3.114  &3.471                            \\
			$\rm ^{33}Mg$   &0.6207 &0.6636&0.2662&0.2390 &3.138  &3.558                            \\
			$\rm ^{34}Mg$   &0.6385 &0.6849&0.2582&0.2331 &3.187  &3.642                            \\
			$\rm ^{35}Mg$   &0.6461 &0.6968&0.2549&0.2302 &3.208  &3.701                            \\
			$\rm ^{36}Mg$   &0.6540 &0.7089&0.2516&0.2271 &3.230  &3.759                            \\
			$\rm ^{37}Mg$   &0.6593 &0.7364&0.2494&0.2191 &3.245  &3.857                            \\
			$\rm ^{38}Mg$   &0.6643 &0.7375&0.2474&0.2191 &3.259  &3.885                            \\
			
		\end{tabular}
	\end{ruledtabular}
\end{table*} 

\begin{table*}
	\caption{The diffuseness parameter for 2pF neutron density distribution, $a_{n}$, and the oscillator constant for SDHO neutron density distribution, $\rm \alpha_n^2$ are obtained from the fitting of experimental reaction cross sections, $\sigma_{R}^{exp}$, on a $^{12}\rm C$ target at 240 MeV/nucleon (given in the last column); the corresponding parameters for proton density distribution ($a_{p}$, $\rm \alpha_p^2$) are the same as obtained in Table \ref{tab1}. $r_{n}$ gives the neutron radius and takes care of the finite size of the nucleon. }
	\renewcommand{\tabcolsep}{0.14mm}
	\renewcommand{\arraystretch}{1.2}
	\label{tab2}
	\begin{ruledtabular}
		\begin{tabular}{ccccccccccccc}
			Projectile & \multicolumn{2}{c}{$\rm 2pF$} &\multicolumn{2}{c}{$\rm SDHO $}&$\sigma_{R}^{exp}$ \cite{19}\\
			\cline{2-3} \cline{4-5} 
			& $a_{n}(fm) $ &$r_{n}(fm)$ & $\alpha^{2}_{n}$$(fm^{-2})$ &$r_{n}(fm) $  &$(mb) $  \\
			\hline
			$\rm ^{24}Mg$   &0.5890&3.0529&0.2573&3.1815 &1172.9$\pm$25.7                                        \\
			$\rm ^{25}Mg$   &0.5623&3.0240&0.2721&3.1329 &1169.5$\pm$14.5                            \\
			$\rm ^{26}Mg$   &0.5463&3.0235&0.2798&3.1224 &1176.3$\pm$11.2                              \\
			$\rm ^{27}Mg$   &0.5085 &2.9711&0.2992&3.0469 &1187.5$\pm$8.5                            \\
			$\rm ^{28}Mg$   &0.5605 &3.1347&0.2722&3.2198 &1244.1$\pm$15.4                              \\  
			$\rm ^{29}Mg$   &0.5465 &3.1366&0.2705&3.2518 &1262.9$\pm$8.6                             \\
			$\rm ^{30}Mg$   &0.5210 &3.1120&0.2812&3.2086 &1263.8$\pm$10.3                             \\
			$\rm ^{31}Mg$   &0.5910 &3.3132&0.2488&3.4289 &1328.9$\pm$9.5                              \\
			$\rm ^{32}Mg$   &0.5998 &3.3670&0.2434&3.4835 &1356.4$\pm$6.8                              \\
			$\rm ^{33}Mg$   &0.6281 &3.4679&0.2355&3.5845 &1398.5$\pm$10.3                              \\
			$\rm ^{34}Mg$   &0.6370 &3.5198&0.2331&3.6420 &1432.2$\pm$15.8                             \\
			$\rm ^{35}Mg$   &0.6212 &3.5098&0.2390&3.6319 &1441.0$\pm$11.0                             \\
			$\rm ^{36}Mg$   &0.6335 &3.5683&0.2358&3.6883 &1471.0$\pm$0.0                            \\
			$\rm ^{37}Mg$   &0.6798 &3.7109&0.2191&3.8569 &1538.0$\pm$13.0                            \\
			$\rm ^{38}Mg$   &0.6560 &3.6780&0.2268&3.8182 &1535.0$\pm$19.0                            \\			
		\end{tabular}
	\end{ruledtabular}
\end{table*}
\begin{table*}
	\caption{In the core+n approach, the diffuseness parameter for 2pF neutron density distribution, $a_{n}$, and the oscillator constant for SDHO neutron density distribution, $\rm \alpha_n^2$, reproduce the corresponding neutron radii, as obtained in Table \ref{tab2}. $r_{n}^{core}$ represents the neutron radius of the core nucleus. The ($\Delta r_{n})\%$ gives the percentage difference between $r_{n}^{core}$ and the neutron radius of the corresponding nucleus in Table \ref{tab2}. The last column gives the projectile one-neutron separation energy.}
	\renewcommand{\tabcolsep}{0.14mm}
	\renewcommand{\arraystretch}{1.2}
	\label{tab3}
	\begin{ruledtabular} 
		\begin{tabular}{ccccccccccccc}
			Projectile &core+n& \multicolumn{3}{c}{$\rm 2pF (core)$} &\multicolumn{3}{c}{$\rm SDHO (core) $}&$S_{n}$ \cite{28}\\
			\cline{3-5} \cline{6-8} 
			&& $a_{n}(fm) $ &$r_{n}^{core}(fm)$ &$(\Delta r_{n})\%$& $\alpha^{2}_{n}$$(fm^{-2})$ &$r_{n}^{core}(fm) $&$(\Delta r_{n})\%$  &$(MeV) $  \\
			\hline
			$\rm ^{25}Mg$    &$\rm ^{24}Mg$+n&0.6075&3.1025&1.6           &0.2591&3.1703&0.4&7.33                             \\     
			$\rm ^{26}Mg$    &$\rm ^{25}Mg$+n&0.5942&3.1070&2.7           &0.2646&3.1746&1.3&11.09                             \\
			$\rm ^{27}Mg$   &$\rm ^{26}Mg$+n&0.5588&3.0549&1.0           &0.2896&3.0694&1.6&6.44                             \\
			$\rm ^{28}Mg$   &$\rm ^{27}Mg$+n&0.6031&3.2062&7.3           &0.2612&3.2611&7.0&8.50                             \\
			$\rm ^{29}Mg$   &$\rm ^{28}Mg$+n&0.5783&3.1790&1.4           &0.2676&3.2471&0.8&3.66                             \\
			$\rm ^{30}Mg$   &$\rm ^{29}Mg$+n&0.5599&3.1691&1.0           &0.2736&3.2338&0.6&6.35                             \\
			$\rm ^{31}Mg$    &$\rm ^{30}Mg$+n&0.6136&3.3369&7.2           &0.2504&3.3997&5.9&2.31                             \\     
			$\rm ^{32}Mg$    &$\rm ^{31}Mg$+n&0.6347&3.4229&3.3           &0.2373&3.5110&2.4&5.78                             \\
			$\rm ^{33}Mg$    &$\rm ^{32}Mg$+n&0.6502&3.4937&3.7           &0.2326&3.5601&2.2&2.28                         \\
			$\rm ^{34}Mg$    &$\rm ^{33}Mg$+n&0.6682&3.5699&2.9           &0.2277&3.6431&1.6&4.71                             \\
			$\rm ^{35}Mg$    &$\rm ^{34}Mg$+n&0.6155&3.4666&1.5           &0.2498&3.5184&3.4&0.75                            \\
			$\rm ^{36}Mg$    &$\rm ^{35}Mg$+n&0.6604&3.6071&2.7           &0.2308&3.6944&1.7&3.33                           \\
			$\rm ^{37}Mg$    &$\rm ^{36}Mg$+n&0.6239&3.5449&0.7           &0.2458&3.6187&1.9&0.24                            \\
			$\rm ^{38}Mg$    &$\rm ^{37}Mg$+n&0.6776&3.7054&0.2           &0.2245&3.8112&1.2&2.21                            \\     	
		\end{tabular}
	\end{ruledtabular}
\end{table*} 
\begin{table*}
	\caption{In the core+n approach, the diffuseness parameter for neutron density distribution, $a_{n}$, and the oscillator constant for neutron density distribution, $\rm \alpha_n^2$, given in Table \ref{tab2}, provides the projectile neutron radius $r_{n}$; the corresponding reaction cross section on a $^{12}\rm C$ target at 240 MeV/nucleon is shown by $\sigma_{R}$. $(\Delta r_{n})\%$ represents the percentage difference between $r_{n}$ in this table and the corresponding $r_{n}$ in Table \ref{tab2}. $(\Delta \sigma_{R})\%$ is the percentage difference between $\sigma_{R}$ in this table and the central value of the corresponding experimental reaction cross section ($\sigma_{R}^{exp}$), given in Table \ref{tab2}.}
	\renewcommand{\tabcolsep}{0.14mm}
	\renewcommand{\arraystretch}{1.2}
	\label{tab4}
	\begin{ruledtabular}
		\begin{tabular}{ccccccccccccc}
			Projectile &core+n& \multicolumn{4}{c}{$\rm 2pF$} &\multicolumn{4}{c}{$\rm SDHO $}\\
			\cline{3-6} \cline{7-10} 
			&& $r_{n}(fm) $ &$(\Delta r_{n})\%$&$\sigma_{R}(mb)$&$(\Delta \sigma_{R})\%$ & $r_{n}(fm)$ &$(\Delta r_{n})\% $&$\sigma_{R}(mb)$&$(\Delta \sigma_{R} )\%$   \\
			\hline
			
			$\rm ^{25}Mg$  &$\rm ^{24}Mg$+n &2.9764&1.6&1164.7&0.4&3.1437&0.3&1175.5&0.5                            \\
			$\rm ^{26}Mg$   &$\rm ^{25}Mg$+n&2.9433&2.7&1164.7&0.9&3.0800&1.3&1170.4&0.5                              \\
			$\rm ^{27}Mg$   &$\rm ^{26}Mg$+n&2.9609 &1.0&1193.7&0.5&3.0976&1.6&1198.8&0.9                           \\
			$\rm ^{28}Mg$   &$\rm ^{27}Mg$+n&2.9066 &7.3&1196.8&3.8&3.0135&6.4&1198.1&3.7                              \\  
			$\rm ^{29}Mg$  &$\rm ^{28}Mg$+n &3.0936 &1.4&1250.4&0.9&3.2255&0.3&1260.5&0.2                             \\
			$\rm ^{30}Mg$   &$\rm ^{29}Mg$+n&3.0805 &1.0&1260.3&0.3&3.2263&0.5&1270.6&0.5                             \\
			$\rm ^{31}Mg$  &$\rm ^{30}Mg$+n &3.0944 &6.0&1270.1&4.4&3.2454&5.3&1286.0&3.2                              \\
			$\rm ^{32}Mg$  &$\rm ^{31}Mg$+n &3.2599 &3.2&1332.6&1.7&3.4037&2.3&1338.0&1.4                              \\
			$\rm ^{33}Mg$  &$\rm ^{32}Mg$+n &3.3444 &3.6&1365.1&2.4&3.5073&2.2&1380.5&1.3                               \\
			$\rm ^{34}Mg$   &$\rm ^{33}Mg$+n&3.4201 &2.8&1417.1&1.1&3.5646&2.1&1411.3&1.5                             \\
			$\rm ^{35}Mg$   &$\rm ^{34}Mg$+n&3.5614 &1.5&1474.8&2.3&3.7502&3.2&1489.1&3.3                             \\
			$\rm ^{36}Mg$  &$\rm ^{35}Mg$+n &3.4731 &3.7&1444.8&1.8&3.6262&1.78&1454.9&1.1                            \\
			$\rm ^{37}Mg$  &$\rm ^{36}Mg$+n &3.7331 &0.6&1482.4&3.6&3.9272&1.8&1593.0&3.5                             \\
			$\rm ^{38}Mg$  &$\rm ^{37}Mg$+n &3.6838 &0.2&1547.9&0.8&3.8631&1.1&1553.9&1.2                             \\			
		\end{tabular}
	\end{ruledtabular}
\end{table*} 

\begin{figure}
	\begin{center}
		
		\includegraphics[height=6.8cm, width=8.9cm]{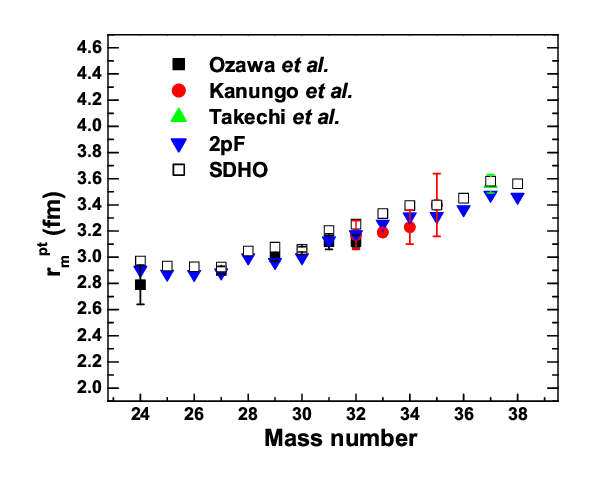}
		\caption{The point matter radii corresponding to our deduced 2pF and SDHO neutron radii (Table \ref{tab2}) and DRHBc proton radii \cite{20}. The results of other works are presented for comparison. The values denoted by Ozawa \textit{et al.}, Kanungo \textit{et al.}, and Takechi \textit{et al.} are taken from Refs. \cite{29}, \cite{30}, and \cite{19}, respectively. }  
		\label{fig4}
	\end{center}
\end{figure}
\begin{figure}
	\begin{center}
		\includegraphics[height=7.8cm, width=8.9cm]{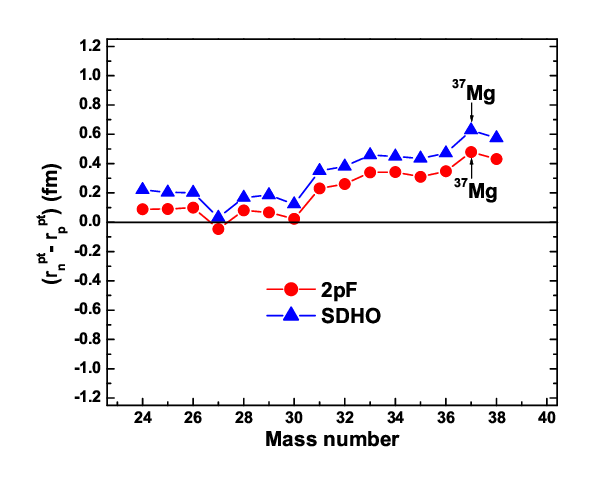}
		\caption{The neutron skin thickness as a function of mass number for $^{24-38}$\rm Mg isotopes, using the point proton \cite{20} and point neutron (Table \ref{tab2}) radii.}  
		\label{fig5}
	\end{center}
\end{figure}
\begin{figure*}
	\begin{center}
		\includegraphics[height=8.0cm, width=16cm]{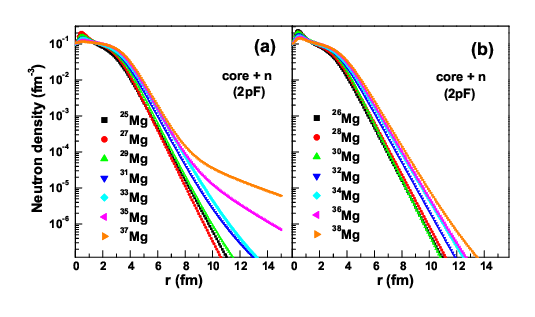}
		\caption{The 2pF neutron density distributions, obtained using the core+n description of $^{25-38}$\rm Mg isotopes; the diffuseness parameter for neutron density distribution $a_{n}$ corresponds to the one obtained in Table \ref{tab3}.  } 
		\label{fig6}
	\end{center}
\end{figure*}
\begin{figure*}
	\begin{center}
		\includegraphics[height=8.0cm, width=16cm]{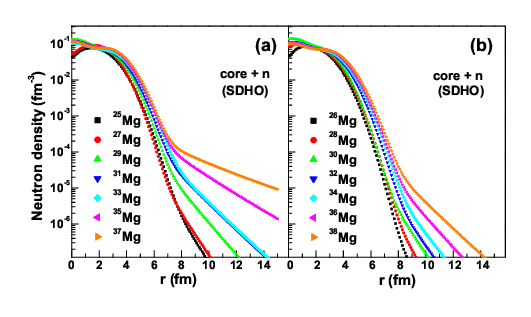}
		\caption{The SDHO neutron density distributions, obtained using the core+n description of $^{25-38}$\rm Mg isotopes; the oscillator constant for neutron density distribution $\rm \alpha_n^2$ corresponds to the one obtained in Table \ref{tab3}.} 
		\label{fig7}
	\end{center}
\end{figure*}

\subsection{Extracting neutron radii of $\textsuperscript{24-38}\rm \text{Mg}$ isotopes by analyzing their reaction cross sections on $\textsuperscript{12}\rm \text{C}$ at 240 MeV/nucleon  }
As we know, the present source for predicting the charge radii of exotic nuclei is the use of their charge-changing cross sestion (CCCS) measurements. Unfortunately the CCCS data for isotopic chain of \rm Mg are not available, one has to involve some model-dependent estimates for the charge radii of \rm Mg isotopes. It is in this context that we rely on the use of DRHBc calculations \cite{20} for charge radii of \rm Mg isotopes for onward calculations. Thus, taking the DRHBc proton radii, we, in the next step, extract the neutron radii of $^{24-38}$\rm Mg isotopes, by varying the oscillator constant ($\rm \alpha^2$) in SDHO density and the diffuseness parameter ($a$) in 2pf density, that fit the experimental values of reaction cross-section for $^{24-38}$\rm Mg isotopes on $^{12}$\rm C target at 240 MeV/nucleon. The extracted neutron radii of $^{24-38}$\rm Mg isotopes are given in Table \ref{tab2}, and the corresponding (point) matter radii ($r_m^{pt}$) are shown in Fig. \ref{fig4}. It is found that our extracted matter radii agree fairly well with those obtained in some earlier calculations \cite{19,29,30}. Fig. \ref{fig5} depicts our deduced neutron skin thickness, $(r_{n}^{pt}-r_{p}^{pt})$, for \rm Mg isotopes as a function of the mass number. In this figure, we notice that although neutron skin thickness follows a similar trend for both the densities, SDHO and 2pF, the absolute values differ from each other; this difference can be related to Fig. \ref{fig1}, which has led to different values of neutron radius, $r_{n}$, (Table \ref{tab2}) for \rm Mg isotopes in order to fit the $\sigma_{R}$ data. However, it should be noted that within the isotopic chain of a \rm Mg element, the nucleus $^{37}$\rm Mg corresponds to the largest neutron skin thickness, which when added to its lowest one-neutron separation energy ($s_{n}=0.240 MeV$ \cite{28}) and a sudden rise in its reaction cross section (Fig. \ref{fig1}) suggests that $^{37}$\rm Mg exhibits one-neutron halo-like structure; this result supports some earlier works by Takechi \textit{et al.} \cite{19} and Sharma \textit{et al.} \cite{31}. For rest of the \rm Mg isotopes, having higher values of $s_{n}$, the order of the thickness of neutron surface can be assessed directly from the measure of $(r_{n}^{pt}-r_{p}^{pt})$.
\subsection{Understanding the asymptotic behavior (spread) of neutron distributions in Mg isotopes: A semi-phenomenological approach }
\begin{figure}
	\begin{center}
		\includegraphics[height=7.9cm, width=8.9cm]{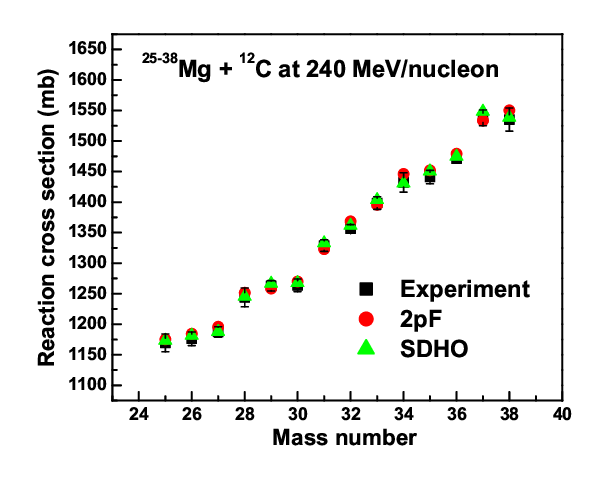}
		\caption{The reaction cross sections of $^{25-38}$\rm Mg isotopes on a $^{12}$\rm C target at 240 MeV/nucleon, using the core+n description for the projectiles. The diffuseness parameter for neutron density distribution, $a_{n}$, and the oscillator constant for neutron density distribution, $\rm \alpha_n^2$, are the same as obtained in Table \ref{tab3}; the corresponding parameters for proton density distribution are taken from Table \ref{tab1}. Filled circles and triangles correspond, respectively, to 2pF and SDHO densities. The experimental data (filled squares) are taken from Ref. \cite{19}. } 
		\label{fig8}
	\end{center}
\end{figure}

The correct description of nucleon wave function plays a crucial role in the theoretical calculations specially for observables of the scattering and reaction processes. Therefore, it seems important to understand the expected asymptotic behavior of the nucleon wave function. Many years ago, Gambhir and Patil \cite{32,33} had proposed an analytic expression for the nucleon densities which incorporates two basic physical requirements: (a) The asymptotic behavior (r $\rightarrow$ $\infty$) and (b) the behavior near the center (r $\rightarrow$ 0). The expression involves the nucleon separation energies ($S_{p}$ and $S_{n}$) and the charge radii as input to get the proton and neutron distributions. It has been shown \cite{32,33} that their density works well for providing the proton and neutron densities near the $\beta$-stability line. For loosely bound nuclei, Bhagwat \textit{et al.} \cite{34} have used a density which contains explicitly an additional term that allows to understand the spread of the neutron density in such nuclei. For example, for a neutron rich (loosely bound) nucleus ($N,Z$), the neutron density is expressed as
\begin{equation}
\rho_{n}(r)=\rho_{core}(r) + \rho_{tail}(r).
\label {eq21}	             	
\end{equation} 
As mentioned in Ref. \cite{34}, both the parts $\rho_{core}$ and $\rho_{tail}$ should have their correct asymptotic behavior; the $\rho_{core}$ in Ref. \cite{34} involves the neutron separation energy for the core nucleus ($N_{core},Z$), and the neutron tail part $\rho_{tail}$ was taken as \cite{34}
\begin{equation}
\rho_{tail}=N_{0}\left(\frac{r^{2}}{(r^{2}+R^{2})^{2}}\right)e^{-r/t_{n}},
\label {eq22}	             	
\end{equation}
where
\begin{equation}
t_{n}=\frac{\hbar}{2(2mS_{n})^{1/2}},
\label {eq23}	             	
\end{equation}
with $S_{n}$ denoting the neutron separation energy of the nucleus ($N,Z$), $m$ is the nucleon mass, the parameter $R$ corresponds to the core nucleus ($N_{core},Z$), which is responsible to reproduce the experimental charge radius of the core nucleus, and $N_{0}$ is fixed by requiring that $\rho_{tail}$ gives the correct number of neutrons in the tail. Here, it should be made clear that number of neutrons in the tail depends upon whether $S_{n}<S_{2n}$ ($S_{2n}$ is the two-neutron separation energy) or $S_{n}>S_{2n}$ for a given nucleus. For $S_{n}<S_{2n}$, $\rho_{tail}$ involves only one neutron, whereas for $S_{n}>S_{2n}$, the corresponding number of neutrons is two. The density given in Eq. (\ref{eq21}) has also been used \cite{34} to calculate the reaction cross sections for loosely bound exotic nuclei (in the neutron-halo region) having very small neutron separation energy. The results are found to be consistent with the experimental values and the calculated  neutron densities are also in good agreement with the experimentally deduced densities (where available). To move further, let us note from Eq. (\ref{eq22}) that the extent of spread in neutron distribution is inversely proportional to the neutron separation energy. In other words, it means that the spread of neutron density from the center of the nucleus is controlled by the neutron separation energy of the nucleus under consideration. We, therefore, conjecture that the density (Eq. (\ref{eq21})) could be used for stable as well as unstable nuclei to predict the expected asymptotic behavior of neutron distribution. It is in this spirit that, in the present work, we have used Eq. (\ref{eq21}) to get the neutron distribution for ($N>Z$) \rm Mg isotopes. Involving the SDHO and 2pF densities for the core nucleus ($N_{core},Z$), the corresponding densities for the nucleus ($N,Z$) (equivalent to Eqs. (\ref{eq21}) and (\ref{eq22})) take the following forms
\begin{equation}
\rho_{n}^{SDHO}(r)=\rho_{core}^{SDHO}(r)+N_{0}\left(\frac{r^{2}}{(r^{2}+\alpha_{n}^{-2})^{2}}\right)e^{-r/t_{n}},
\label {eq24}	             	
\end{equation}
\begin{equation}
\rho_{n}^{2pF}(r)=\rho_{core}^{2pF}(r)+N_{0}\left(\frac{r^{2}}{(r^{2}+a_{n}^{2})^{2}}\right)e^{-r/t_{n}}.
\label {eq25}	             	
\end{equation}
Here, it should be noted that, like in Ref. \cite{34}, both the core and tail parts have the same parameter; the use of SDHO and 2pF (core) densities involve $\alpha_{n}^{2}$ and $a_{n}$ as the respective parameters. However, the only difference between our approach and the one used in Ref. \cite{34} is that we have treated the $\alpha_{n}^{2}$ ($a_{n}$) in the case of SDHO (2PF) density as a variable parameter that restricts to reproduce our extracted neutron radii as predicted by analyzing the experimental reaction cross sections (Table \ref{tab2}), whereas, in Ref. \cite{34}, the parameter (R) appearing in the expressions for $\rho_{core}$ and $\rho_{tail}$ (Eq. (\ref{eq21})) is fixed from the charge radius of the core nucleus ($N_{core},Z$), and the neutron radius as well as the reaction cross section are then predicted for the given nucleus ($N,Z$). In our opinion, if we are concerned with the neutron distribution of a given nucleus ($N,Z$), it is advisable to consider the neutron distribution of the core nucleus ($N_{core},Z$) in order to get the result for neutron distribution of the nucleus under consideration; this approach goes against taking the charge radius of the core ($N_{core},Z$) for predicting the neutron distribution of the nucleus ($N,Z$) \cite{34}. It may also be mentioned at this stage that, in this subsection, all the considered $^{25-38}$\rm Mg isotopes correspond to $S_{n}<S_{2n}$, we shall use the core+n description for these isotopes in the onwards calculations. To move further, let us now vary (i) the parameter $\alpha_{n}^{2}$ in Eq. (\ref{eq24}) for SDHO density, and (ii) the parameter $a_{n}$ in Eq. (\ref{eq25}) for 2pF density in order to get our extracted neutron radii (Table \ref{tab2}) for ($N>Z$) \rm Mg isotopes. The respective parameters for 2pF and SDHO densities are given in Table \ref{tab3}. Table \ref{tab3} also presents the neutron radius of the core nucleus in 2pF and SDHO calculations for a given ($N>Z$) nucleus. $(\Delta r_{n})\%$ shows the percentage difference between the neutron radii of the core nucleus (Table \ref{tab3}) and the corresponding nucleus in Table \ref{tab2}. The values of $(\Delta r_{n})\%$ in both the cases (2pF and SDHO) suggest that if we need to translate our neutron radii (Table \ref{tab2}) into core+n(tail) description of ($N>Z$) \rm Mg isotopes, one demands different neutron radius of the core nucleus (Table \ref{tab3}) as compared to that (Table \ref{tab2}) if studied in the free state. To understand this difference, let us visualize the core+n description of a given nucleus ($N,Z$) as equivalent to the one adding a neutron to the core nucleus ($N_{core},Z$) (in the free state) for getting the nucleus ($N,Z$). It might happen that the interaction between this additional neutron and the core nucleus (now in the bound state of the nucleus ($N,Z$)) does not allow to have the free state of the core nucleus as intact, thereby leading to different neutron radius of the core in order to reproduce our extracted neutron radius (Table \ref{tab2}) in the core+n description of the given ($N>Z$) nucleus. Here, it should be added that a similar argument relating the interaction between the additional neutron and the core nucleus may be applicable to the proton radius of the core nucleus also. Due to this, the proton radius of the core nucleus may differ from the corresponding nucleus in the free state. However, it has been verified (see results, in the next paragraph, presented in Fig. \ref{fig8}), that the core+n description of ($N>Z$) \rm Mg isotopes, involving the DRHBc proton radii \cite{20} along with our extracted neutron radii (Table \ref{tab2}), provides a quite satisfactory explanation of the experimental reaction cross sections for $^{25-38}$\rm Mg isotopes from $^{12}$\rm C at 240 MeV/nucleon. Keeping this in view, we, therefore, assumed similar value of the proton radius for the core nucleus ($N_{core},Z$) whether it is in the free state or considered in the nucleus ($N,Z$).

Figs. \ref{fig6} and \ref{fig7} show, respectively, the predicted asymptotic behavior (spread) of 2pF and SDHO neutron density distributions obtained using the core+n description for ($N>Z$) \rm Mg isotopes. It is noticed that $^{37}$\rm Mg shows the maximum spread far out from the center as compared to all other \rm Mg isotopes. This result supplements the known one-neutron halo structure of $^{37}$\rm Mg nucleus. To verify the utility of our extracted neutron radii for \rm Mg isotopes (Table \ref{tab2}), and to establish credibility for the use of core+n description for $^{25-38}$\rm Mg isotopes, we have revisited the reaction cross sections of $^{25-38}$\rm Mg from a $^{12}$\rm C target at 240 MeV/nucleon. The results of such calculations are presented in Fig. \ref{fig8}. It is found that the core+n description of $^{25-38}$\rm Mg isotopes, along with our neutron radii (Table \ref{tab2}), works well and one is able to provide fairly good account of the experimental data using both the 2pF and SDHO densities.

Now, out of curiosity, the core+n description has also been used to find the neutron radii of $^{25-38}$\rm Mg isotopes and predict their reaction cross sections on a $^{12}$\rm C target at 240 MeV/nucleon, involving the (free state) parameter $\alpha_{n}^{2}$ in Eq. (\ref{eq24}) ($a_{n}$ in Eq. (\ref{eq25})) given in Table \ref{tab2}. In these calculations, our motive is to see how far the neutron radii and the reaction cross sections, obtained in this way, agree, respectively, with the corresponding extracted values (Table \ref{tab2}) and the experiment \cite{19}. The results of such calculations are presented in Table \ref{tab4}. It is found that the results on neutron radii and reaction cross sections, respectively, agree well with the corresponding extracted (Table \ref{tab2}) and experimental \cite{19} values; only a few cases show a difference of $\leq 7\%$ ($\leq 4\%$) in the case of neutron radii (reaction cross sections). This suggests that the present core+n approach could be satisfactorily used to assess the credibility of the given neutron radii (involving the free state behavior of the core nucleus) in predicting the reaction cross sections for the isotopes of other elements. Moreover, the neutron radius  of the core nucleus (if available) may find its place to predict the neutron radii and reaction cross sections for the core+n as well as core+2n nuclei, where measurements are scarce.
 \begin{table*}
 	\caption{In the core+2n approach, $r_{n}$ and $r_{m}$ represent the predicted neutron and matter radii of $\rm ^{40}Mg$. $\sigma_{R}$ gives the predicted reaction cross section of $\rm ^{40}Mg$ on a $\rm ^{12}C$ target at 240 and 1000 MeV/nucleon. The last column represents the predictions by Singh \textit{et al.}\cite{21}.  $r_{n}$ and $r_{m}$ take care of the finite size of the nucleon. }
 	\renewcommand{\tabcolsep}{0.14mm}
 	\renewcommand{\arraystretch}{1.2}
 	\label{tab5}
 	\begin{ruledtabular}
 		\begin{tabular}{ccccccccccccc}
 			Projectile &Energy&core+2n&\multicolumn{3}{c}{$\rm 2pF$} &\multicolumn{3}{c}{$\rm SDHO$}&\multicolumn{2}{c}{$\rm Singh$ \textit{et al.} \cite{21}}\\
 			\cline{4-6} \cline{7-9} \cline{10-11} 
 			&(MeV/nucleon)&& $r_{n}(fm) $ &$r_{m}(fm)$&$\sigma_{R}(mb)$ &$r_{n}(fm)$& $r_{m}$(fm) &$\sigma_{R}(mb)$&$r_{m}(fm)$  &$\sigma_{R}(mb)$  \\
 			\hline
 			$\rm ^{40}Mg$&    &$\rm ^{38}Mg$+2n&3.8472&3.6873&           &4.1076&3.8789&      &3.8237&                             \\     
 			&240 &                &      &      &1646.9     &      &      &1688.1&   &1601-1807                             \\
 			&1000&                &      &      &1742.4    &        &      &1783.1&   &1695-1944                             \\
 			
 		\end{tabular}
 	\end{ruledtabular}
 \end{table*}  
\subsection{Predictions for matter radius of $\textsuperscript{40}\rm \text{Mg}$ and its reaction cross sections from $\textsuperscript{12}\rm \text{C}$ at 240 and 1000 MeV/nucleon}

\begin{figure}
 	 	\begin{center}
 		\includegraphics[height=7.2cm, width=8.9cm]{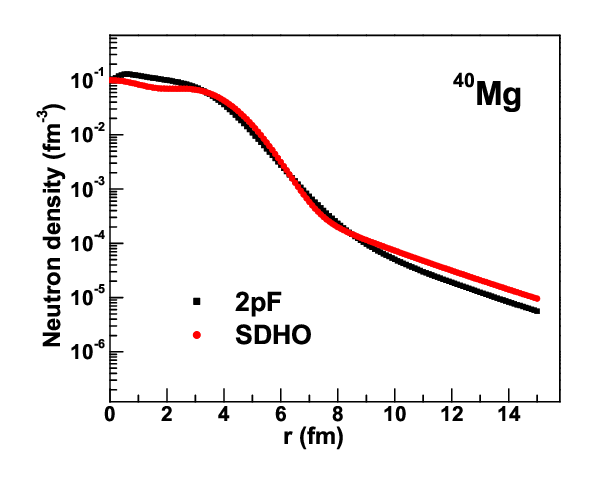}
 		\caption{The 2pF (filled squares) and SDHO (filled circles) neutron density distributions in $^{40}$\rm Mg nucleus, obtained using its core+2n description. For the core nucleus ($^{38}$\rm Mg), the diffuseness parameter in 2pF neutron density distribution, $a_{n}$, and the oscillator constant in SDHO neutron density distribution, $\rm \alpha_n^2$, are the same as obtained in Table \ref{tab2}; the corresponding parameters for proton density distribution are taken from Table \ref{tab1}.} 
 		\label{fig9}
 	\end{center}
 \end{figure}

As mentioned above, the neutron radius of the core nucleus (in the free state) could be used to predict the neutron radii and reaction cross sections for the core+n and also core+2n nuclei, where experimental data are not available. Unfortunately, the $^{40}$\rm Mg nucleus lacks in its reaction cross section on $^{12}$\rm C at any incident energy. Moreover, we found that the one-neutron separation energy ($S_{n}$=1.30 MeV \cite{28}) of $^{40}$\rm Mg is larger than its two-neutron separation energy ($S_{2n}$=0.67 MeV \cite{28}) . Therefore, one may use core+2n description of $^{40}$\rm Mg, taking $^{38}$\rm Mg as a core nucleus, for which our extracted neutron radius (Table \ref{tab2}) has been used. The results of such calculations, using SDHO (Eq. (\ref{eq24})) and 2pF (Eq. (\ref{eq25})) densities are presented in Table \ref{tab5}. In this table, we have shown our predictions for neutron and matter radii of $^{40}$\rm Mg, and its reaction cross sections on a $^{12}$\rm C target at 240 and 1000 MeV/nucleon. It is found that our predictions support the results obtained using a three-body model ($^{38}$\rm Mg+n+n) for $^{40}$\rm Mg with effective n-n and $^{38}$\rm Mg+n interactions \cite{21} (also given in Table \ref{tab5}). Moreover, we notice that the predicted values of the reaction cross sections for $^{40}$\rm Mg show significant increase with respect to the experimental trend of $\sigma_{R}$ in the lower-A \rm Mg isotopes at both 240 MeV/nucleon \cite{19} and 1000 MeV/nucleon \cite{29,30}. Further, we find that the neutron radius ($r_{n}$) of $^{40}$\rm Mg also shows enhancement with respect to the $r_{n}$ of its neighboring nucleus $^{38}$\rm Mg. Thus, we arrive at the result that the basic requirements, the lowest two-neutron separation energy, and enhanced values of neutron radius and $\sigma_{R}$, support the two-neutron halo structure of $^{40}$\rm Mg, a result similar to that observed by $\rm Singh$ \textit{et al.}\cite{21}. Finally, Fig. \ref{fig9} depicts our predicted 2pF and SDHO neutron density distributions of $^{40}$\rm Mg, employing its core+2n description. 
                              
\section{Summary and Conclusions}
\label{sec4}

Working within the framework of Glauber model, and involving the proton radii of \rm Mg isotopes, as calculated using the deformed relativistic Hartree-Bogoliubov theory in continuum (DRHBc) \cite{20}, we have extracted the neutron radii of $^{24-38}$\rm Mg isotopes by analyzing their reaction cross sections ($\sigma_{R}$) on a $^{12}$\rm C target at 240 MeV/nucleon. The calculations use (i) descriptions of nuclei in terms of the Slater determinant involving harmonic oscillator single-particle wave functions (SDHO), and (ii) two-parameter Fermi (2pF) shape of density distribution, with the aim to see the density dependence of neutron skin in $^{24-38}$\rm Mg isotopes. To understand the asymptotic behavior (spread) of neutron distribution in $^{25-38}$\rm Mg isotopes ($S_{n}<S_{2n}$), the core+n description has been employed. The core+n has been treated semi-phenomenologically, and is subjected to reproduce the same neutron radius of the given \rm Mg isotope, as we predicted, in this work, from the study of $\sigma_{R}$ data. To test the core+n description, we have recalculated the reaction cross sections of $^{25-38}$\rm Mg isotopes. The results are found to provide a good agreement with the experimental values. Moreover, the core+n neutron distributions clearly demonstrate the one-neutron halo structure of $^{37}$\rm Mg. Encouraged by the successful use of core+n description of $^{25-38}$\rm Mg isotopes, we have extended our study to predict the neutron radius of $^{40}$\rm Mg ($S_{n}>S_{2n}$), and also its $\sigma_{R}$ on a $^{12}$\rm C target at 240 and 1000 MeV/nucleon. For this, we have represented $^{40}$\rm Mg as $^{38}$\rm Mg+2n. It is found that the values of our extracted neutron radius and $\sigma_{R}$ agree well with those obtained by $\rm Singh$ \textit{et al.} \cite{21}, and also support their prediction that $^{40}$\rm Mg exhibits two-neutron halo-like structure.   

In conclusion, the present work suggests that the core+n (core+2n) approach seems promising to assess the credibility of the given neutron radii in predicting the reaction cross sections for the isotopes of other elements having $S_{n}<S_{2n}$ ($S_{n}>S_{2n}$). Moreover, the neutron radius of the core nucleus (if available) may find its place to predict the neutron radii and reaction cross sections for the core+n as well as core+2n nuclei, where measurements are scarce.

\section{ Acknowledgments}

G.K acknowledges the UGC-Non Net Fellowship as per UGC letter No.F.19-33/2006/(CU) dated 01/02/2011, and Office Order No. 223/FO/Sch/2010. Z.H. acknowledges the UGC-BSR Research Start-Up-Grant (No.F.30-310/2016(BSR)). M.I. and Z.A.K. acknowledge the Department of Physics, Aligarh Muslim University, Aligarh, India for using the computational facility.

\end{document}